\documentclass[10pt]{article}
\usepackage{amsfonts}
\usepackage{amsmath}
\usepackage{graphicx}
\begin{document}

\title{Two-Photon Airy Disk}
\author{ Hai-Bo Wang, Xin-Bing Song, De-Qin Xu, Jun Xiong, Xiangdong Zhang,
and Kaige Wang\footnote{%
Author to whom correspondence should be addressed. Electric mail:
wangkg@bnu.edu.cn}\\
%EndAName
\it Department of Physics, Applied Optics Beijing Area Major Laboratory, \\
\it Beijing Normal University, Beijing 100875, China}
\maketitle

\begin{abstract}
We report an experimental observation of quantum Airy disk diffraction
pattern using an entangled two-photon source. In contrast to the previous
quantum lithography experiments where the subwavelength diffraction patterns
were observed in the far field limit, we perform the Fraunhofer diffraction
experiment with a convex lens. The experimental result shows that the
two-photon Airy disk is provided with the super-resolution spot, which
surpasses the classical diffraction limit. In particular, the spot size can
be well controlled by the focal length, which adapted to optical
super-focusing. Our experiment can promote potential application of quantum
lithography.
\end{abstract}

Focusing effect of a light beam has potential applications in many areas
such as micro-optical fabrication, high-precision processing,
high-resolution imaging and photolithography, etc. In classical optics, the
minimum size of the focused light spot is governed by the diffraction limit.
Recently, theoretical and experimental studies showed that quantum
interference with a multiphoton entangled state can surpass the classical
diffraction limit\cite{yam,boto,monken,shih1,bj,bd,ed,pe,pee,ko}.
Theoretically, Jacobson et al. \cite{yam} proved that the de Broglie
wavelength of $N$-photon packet is $\lambda /N$, where $\lambda $ and $N$
are the wavelength and number of the constituent photons, respectively. Boto
et al. \cite{boto} proposed a scenario of quantum interferometric optical
lithography with N-photon entangled state to beat the diffraction limit. The
theoretical proposals were then implemented in the experiments using
two-photon entangled state generated by spontaneous parametric
down-conversion(SPDC)\cite{monken,shih1,ed,pee}. In a frequency degenerate
type-II SPDC, a pump photon with frequency $\omega_{p}$ impinges on a $\beta$%
-BaB$_2$O$_4$ (BBO) crystal to generate a pair of entangled photons with the
frequencies $\omega _{s}\approx \omega _{i}=\omega _{p}/2$ in different
polarized modes. When the photon pairs are projected onto a double slit, the
interference-diffraction pattern can be observed by the two-photon
coincidence measurement. The pattern exhibits a smaller spatial modulation
period which beats the classical diffraction limit by a factor of two\cite%
{shih1}. Moreover, it was noticed recently that the two-photon subwavelength
interference for a double-slit can also be realized in a similar way with a
well-designed interferometer driven by a thermal light source\cite{cao}.

In the quantum lithography experiments with a two-photon entangled state,
the resolution improvement of the interference-diffraction pattern exists in
the far-filed limit where Fraunhofer diffraction occurs\cite%
{monken,shih1,ed,pee}. However, the pattern to be observed in the far-field
has been expanded in a large scale, which is generally larger than the
object size. Hence the far-field pattern would not be appropriate for the
technical application of photolithography. On the other hand, our recent work%
\cite{song} showed that there is no net improvement of the transverse
resolution in the near-field Fresnel diffraction, such as two-photon Talbot
self-imaging. Instead, the two-photon subwavelength effect is reflected in
twice the propagation distance for one-photon diffraction.

In this paper we consider two-photon Fraunhofer diffraction using a lens. We
find that in this scheme the resolution of the diffraction pattern is
enhanced to beat the classical diffraction limit, as the same quantum
lithography for the far field case. At the same time, the spot size of the
pattern can be well controlled by the focal length.

Let a two-photon entangled state be
\begin{equation}
|\Psi\rangle=\int d \mathbf{r}_1 d \mathbf{r}_2 C(\mathbf{r}_1,\mathbf{r}%
_2)a^{\dag}_1(\mathbf{r}_1)a^{\dag}_2(\mathbf{r}_2)|0\rangle,  \label{1}
\end{equation}
where $a^{\dag}_j (j=1,2)$ are the photon creation operators for the two
SPDC modes, and $\mathbf{r}_j (j=1,2)$ are their transverse coordinates
across the beam. $C(\mathbf{r}_1,\mathbf{r}_2)\sim\langle0|E^{(+)}_{s1}(%
\mathbf{r}_1)E^{(+)}_{s2}(\mathbf{r}_2)|\Psi\rangle$ characterizes the
two-photon wavepacket with the field operators $E^{(+)}_j (j=s1,s2)$ of mode
$j$ in the source plane. After propagation, the diffraction field in the
observation plane is given by
\begin{equation}
E^{(+)}(\mathbf{r})=\int d \mathbf{r}_0 h(\mathbf{r},\mathbf{r}_0)E^{(+)}_s(%
\mathbf{r}_0),  \label{2}
\end{equation}
where $h(\mathbf{r},\mathbf{r}_0)$ is the impulse response function. So the
two-photon wavepacket in the observation plane is obtained as
\begin{eqnarray}
&&\langle0|E^{(+)}_1(\mathbf{r}_1)E^{(+)}_2(\mathbf{r}_2)|\Psi\rangle  \notag
\\
&&= \int d \mathbf{r}^{\prime }_0 d \mathbf{r}^{\prime \prime }_0 h_1(%
\mathbf{r}_1,\mathbf{r}^{\prime }_0)h_2(\mathbf{r}_2,\mathbf{r}^{\prime
\prime }_0) C(\mathbf{r}^{\prime }_0,\mathbf{r}^{\prime \prime }_0),
\label{3}
\end{eqnarray}
where $h_j$ is the impulse response function for the field mode $j$. The
two-photon coincidence counting rate reads
\begin{eqnarray}
R(\mathbf{r}_1,\mathbf{r}_2)&\propto&\langle\Psi| E^{(-)}_1(\mathbf{r}%
_1)E^{(-)}_2(\mathbf{r}_2)E^{(+)}_2(\mathbf{r}_2)E^{(+)}_1(\mathbf{r}%
_1)|\Psi\rangle  \notag \\
&=&|\langle0|E^{(+)}_1(\mathbf{r}_1)E^{(+)}_2(\mathbf{r}_2)|\Psi\rangle|^2.
\label{4}
\end{eqnarray}

For simplicity, we consider an ideal two-photon entangled state at the
source, which satisfies $C(\mathbf{r}^{\prime }_0,\mathbf{r}^{\prime \prime
}_0)=\delta(\mathbf{r}^{\prime }_0-\mathbf{r}^{\prime \prime }_0)$. Eq.(\ref%
{3}) is written as
\begin{equation}
\langle0|E^{(+)}_1(\mathbf{r}_1)E^{(+)}_2(\mathbf{r}_2)|\Psi\rangle=\int d
\mathbf{r}_0 h_1(\mathbf{r}_1,\mathbf{r}_0)h_2(\mathbf{r}_2,\mathbf{r}_0).
\label{5}
\end{equation}
The impulse response function from the front to the back focal planes of a
thin lens is given by $h_f(\mathbf{r},\mathbf{r}_0)=1/(i\lambda f)\times
\exp [i4\pi f/\lambda]\exp [-i2\pi (\mathbf{r}\cdot \mathbf{r}_0)/(\lambda
f)]$, where $\lambda$ is the wavelength and $f$ is the focal length of the
lens. We put the two-photon source at the front focal plane of the lens and
the detector at the back one. Assuming a mask object of the transmittance
function $P(\mathbf{r}_0)$ (note that $P^2(\mathbf{r}_0)$=$P(\mathbf{r}_0)$)
is placed at the source, we obtain the two-photon coincidence counting rate
\begin{eqnarray}
R(\mathbf{r}_1,\mathbf{r}_2)&\propto& \left\vert \int P(\mathbf{r}_0)\exp
[-i2\pi (\mathbf{r}_1+\mathbf{r}_2)\cdot \mathbf{r}_0/(\lambda f)] d\mathbf{r%
}_0\right\vert^2  \notag \\
&\propto& \left\vert \widetilde{P}[2\pi (\mathbf{r}_1+\mathbf{r}_2)/(\lambda
f)]\right\vert^2,  \label{6}
\end{eqnarray}
where $\widetilde{P}$ is the Fourier transform of function $P$.

Let a circle aperture of radius $a$ be placed at the front focal plane of
the lens, we then calculate the normalized two-photon coincidence counting
distribution for $\mathbf{r}_1=\mathbf{r}_2=\mathbf{r}$ in the back focal
plane
\begin{equation}
R(\mathbf{r})/R(0)=\left\vert \frac{2J_{1}(\frac{2\pi}{\lambda f}2ar)}{\frac{%
2\pi}{\lambda f}2ar}\right\vert^2,  \label{7}
\end{equation}
where $R(\mathbf{r})\equiv R(\mathbf{r},\mathbf{r})$; $J_1$ is the Bessel
function of the first kind and $r=|\mathbf{r}|$. This is the well-known Airy
disk, a focused spot pattern when a plane wave incident on a circle lens or
a diffraction pattern of a circular source in the far-field limit. But the
present Airy disk has a two times resolution of the classical one, and hence
the size of the central spot is reduced to half.

\begin{figure}[htb]
\centering
\includegraphics[width=5cm]{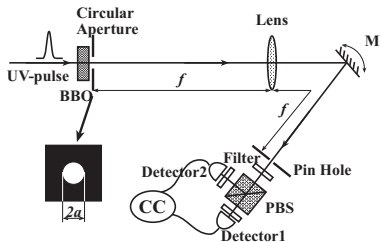}
\caption{Schematic diagram of experimental setup. Photon pairs are generated
by a BBO crystal pumped by a UV pulse laser. A circular aperture closed to
the BBO and a pin hole are placed in the front and back focal planes of a
lens, respectively. PBS is a polarizing beam splitter and M is a mirror.
Coincidence counting is implemented by two detectors.}
\end{figure}

To demonstrate the two-photon Airy disk described above, we perform an
experiment which is sketched in Fig. 1. The UV-pulse beam is provided by a
second-harmonic of a Ti:sapphire femtosecond laser (Mira-900 Coherent Inc.)
with the central wavelength $\lambda=400$nm and repetition rate 76 MHz. The
UV-pulse is used to pump a $5\times5\times2$mm type-II BBO crystal to
generate collinear orthogonally polarized photons via the spontaneous
parametric down-conversion (SPDC) process. A circular aperture is placed
immediately after the crystal and at the front focal plane of the lens. A
pin hole is placed at the back focal plane. After passing through the pin
hole, the pump beam is blocked by a cutoff filter. However, the entangled
photon pairs are separated by a polarizing beam splitter (PBS) and then
detected by two single-photon detectors (Perkin-Elmer SPCM-AQR-14). Both
detectors are preceded by $10$nm bandwidth interference filters centered at
the degenerate wavelength of $800$nm. The pin hole, PBS and the coincidence
circuit work together as a two-photon detector. For the sake of convenience,
instead of moving the two detectors together, we rotate the mirror M to
``scan'' the diffraction pattern across the detector surface\cite{shih1}.
The time window for the coincidence counts is $2$ns.

\begin{figure}[htb]
\centering
\includegraphics[width=5.5cm]{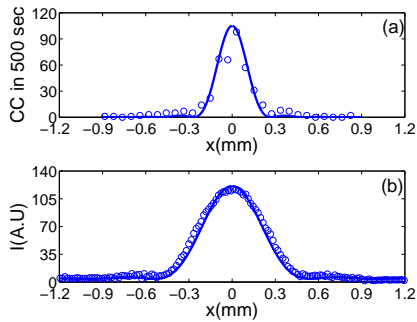}
\caption{Experimental observation of the Airy disk patterns with (a) a
two-photon entangled source, and (b) a classical coherent source, where the
experimental curves in (a) and (b) are recorded by the two-photon
coincidence measurement and the intensity measurement, respectively. The
open circles are the experimental data and the solid lines are the
theoretical curves.}
\end{figure}

In our scheme, the diameter of the circular aperture is $2a=0.9$mm, and the
focal length of the lens is $f=50$cm. The experimental results are shown in
Fig. 2, where the experimental data and theoretical curves are indicated by
open circles and solid lines, respectively. Figure 2(a) shows the two-photon
coincidence counting distribution for the two-photon entangled source. For
comparison, we also use a laser beam with the same wavelength of 800nm to
replace the down-converted beams and impinge on the circular aperture. The
diffraction pattern is recorded by a CCD camera set at the back focal plane
of the lens. The conventional Airy disk has been observed in Fig. 2(b). We
can see that the two-photon Airy disk in Fig. 2(a) has a more subtle central
spot with the half size of that for the classical Airy disk in Fig. 2(b).
For the both cases, the theoretical curves fit well with the experimental
results.

In summary, we have experimentally observed the quantum Airy disk
diffraction pattern with an entangled two-photon source. In contrast to the
previous two-photon quantum lithography experiments where the far-field
diffraction was observed, the present scheme is carried out by a convex lens
to perform the Fraunhofer diffraction. The experimental result has
demonstrated that the two-photon Airy disk can surpass the classical
diffraction limit, achieving the super-resolution focusing effect. Therefore
our experiment is not only interesting from a fundamental point of view, but
also relevant for the quantum lithography application.

This work was supported by the National Natural Science Foundation of China,
Project Nos. 11174038 and 10825416, and the National High Technology
Research and Development Program of China, Project Grant No. 2011AA120102,
and the Fundamental Research Funds for the Central Universities.

\end{document}